\documentclass[osajnl,twocolumn,showpacs,superscriptaddress,10pt]{revtex4-1} 
\usepackage{amsmath,amssymb,graphicx}
\begin{document}

\title{Saturated-absorption spectroscopy revisited: atomic transitions in strong magnetic fields ($>$20~mT) with a micrometer-thin cell}

\author{A.~Sargsyan}
\author{A.~Tonoyan}
\author{R.~Mirzoyan}
\author{D.~Sarkisyan}
\affiliation{Institute for Physical Research, NAS of Armenia, Ashtarak, 0203, Armenia}

\author{A.M.~Wojciechowski}\email{Corresponding author: a.wojciechowski@uj.edu.pl}
\author{A.~Stabrawa}
\author{W.~Gawlik}
\affiliation{Institute of Physics, Jagiellonian University, Reymonta 4, 30-059 Krak\'ow, Poland}

\begin{abstract}
The existence of cross-over resonances makes saturated-absorption spectra very complicated when external magnetic field B is applied. It is demonstrated for the first time that the use of micrometric-thin cells (MTC, $L\approx40\,\mu$m) allows application of SA for quantitative studies of frequency splittings and shifts of the Rb atomic transitions in a wide range of external magnetic fields, from 0.2 up to 6 kG (20-600~mT). We compare the SA spectra obtained with the MTC with those obtained with other techniques, and present applications for optical magnetometry with micrometer spatial resolution and a broadly tunable optical frequency lock.
\end{abstract}



\maketitle 
Saturated absorption (SA) spectroscopy is widely used in laser spectroscopy of atomic and molecular transitions \cite{1, 2, 4}. In this technique, the laser beam is split into a weak probe and a strong pump fields, which are sent to the interaction cell as counter-propagating, overlapping beams. Opposite Doppler shifts and saturation by the pump beam allow the observation of a Doppler-free dip in the probe beam absorption, the so-called velocity selective optical pumping/saturation (VSOP) resonance located at the line center \cite{2}. The line width of the VSOP resonance may be as narrow as the natural width of the transition, which finds applications in realization of frequency references. The situation is more complex when several hyperfine atomic transitions overlap within the Doppler profile, which is the case for the majority of real atomic lines. Multiple unresolved hyperfine transitions result in the formation of the so called crossover (CO) resonances. The COs  are due to atoms that move with a particular non-zero velocity relative to the light beams and may seriously complicate the applicability of the SA spectral reference technique.
In the case of hyperfine levels with a small (10--30 MHz) frequency spacing, the CO resonances may mask the real atomic VSOP resonances, i.e., those due to the atoms moving with zero longitudinal velocity, and hinder or even impede identification of the relevant spectral features.

Several techniques allow for the elimination of crossover resonances in atomic vapors \cite{5,6}.
Similar VSOP resonances have been demonstrated in the transmission spectrum of one dimensional nanometric-thin cell (NTC) filled with alkali metal, with a thickness $L=\lambda$, where $\lambda$ is the transition wavelength, yet the CO lines are absent \cite{7,8}.

In high magnetic fields the atomic optical transitions may dramatically change their frequencies and probabilities, e.g., in the Back-Goudsmit (BG or hyperfine Paschen-Back) regime \cite{9,10}. The related effects were recently studied in the fields up to $B \sim\!\!0.6$~T with a 1x1x1~mm$^3$ Rb-vapor cell embedded in a permanent magnet \cite{11}. Even for that large B-field values, the Doppler-broadened $^{85}$Rb and $^{87}$Rb lines are strongly overlapping. In order to eliminate the Doppler broadening, the SA (and polarization) spectroscopy in the weak/intermediate magnetic fields (up to $\sim100$~G) was implemented to the studies of atomic transitions \cite{13,14} or for calibration of the magnetic field \cite{15}. However, the obtained spectra were rather complicated, primarily due to the presence of strong CO resonances also split into many components \cite{26}.
Thus, the SA spectroscopy in the typical vapor cells is practical only for $B << 200 $~G. In this paper we demonstrate that for magnetic fields $B > 200 $~G, to about 6~kG, the SA spectroscopy can be implemented successfully using 20-50 $\mu$m-thin cells.

The MTC is similar in the design to the extremely thin cells presented in Ref. \cite{7}. The rectangular 20x30~mm$^2$, 2.5~mm-thick sapphire window wafers (chemically resistant to hot alkali metal vapors), polished to $<1$~nm surface roughness, form a vapor cell, wedged by placing
spacers between the windows prior to gluing. MTC is filled with a natural mixture of rubidium isotopes. The wedged gap allows one to study the SA spectra with a variable column thickness by probing various cell regions.

The small thickness of MTC is advantageous for the application of very strong magnetic fields with the use of permanent ring magnets, otherwise not practical for spectroscopy because of strong magnetic field gradients. In MTC, however, the B-field inhomogeneity over the narrow atomic vapor column is lower by a few orders of magnitude than the applied B values. The magnets were mounted onto two nonmagnetic stages with the possibility of  adjusting their distance.

\begin{figure}[tb]
	\includegraphics[width=.8\columnwidth]{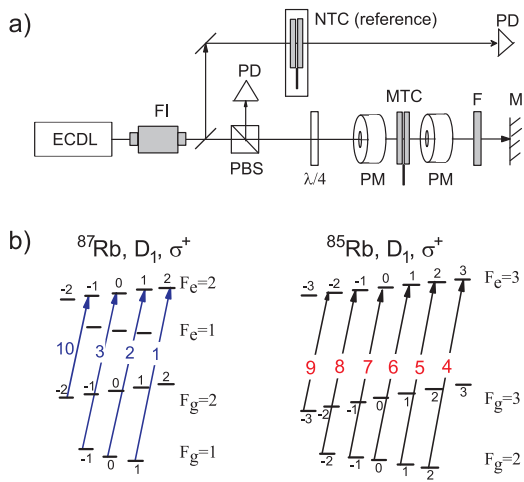}
	\caption{(a) Sketch of the experimental setup. ECDL is the diode laser, FI - Faraday isolator, PM - permanent magnet, $\lambda$/4 - quarter waveplate, PBS - polarization beam splitter, PD - photodetectors, F - filter, M - mirror; (b) Energy level diagram of the $D_1$ line of the $^{87}$Rb (left) and $^{85}$Rb (right) in a magnetic field with the $\sigma^+$ transitions of non-vanishing probability in the BG regime.}
		\label{fig:setup}
\end{figure}

The laser beam from an extended cavity diode laser resonant with the Rb $D_1$ line ($\lambda = 795$~nm, beam diameter of $\sim 1$~mm) was directed through MTC with the vapor column thickness $L = 40\,\mu$m [Fig. \ref{fig:setup}(a)]. To achieve  $\sigma^{+}$ circular polarization of the laser radiation, we used a polarizing beam splitter followed by a $\lambda /4$ plate. After passing the MTC, the beam was retroreflected and the transmission signal was detected by a photodiode and recorded by a digital scope. The MTC was heated to ${\sim} 70 ^{\circ}$C by a hot air to yield atomic number density of the order of $N = 5\!\cdot\! 10^{11}$~cm$^{-3}$. The magnetic field was oriented along the beam propagation direction. A fraction of the laser power was sent to the reference arm with the Rb-filled NTC of a thickness $L= \lambda$ or $\lambda$/2 and the transmission spectrum was used as a $B =0$ frequency reference.

For high magnetic fields $B \gg 200$~G, the only allowed transitions between magnetic sublevels of the hf states for $^{87}$Rb and $^{85}$Rb $D_1$ lines in case of $\sigma^{+}$ polarized excitation are shown in Fig. \ref{fig:setup}(b). Only four (six) atomic transitions for $^{87}$Rb ($^{85}$Rb) remain in the transmission spectrum, while there are twelve (twenty) Zeeman transitions allowed at low fields. This strong reduction of the numbers of atomic transitions for large magnetic field is caused by the effect of decoupling of the total electronic momentum $J$ and the nuclear spin momentum $I$ (the BG effect) in the magnetic field $B > B_0 \equiv A_{\textrm{hfs}}/\mu_B$, where $A_{\textrm{hfs}}$ is the ground-state hyperfine coupling coefficient and $\mu_B$ is the Bohr magneton. Thus, with increasing B majority of Zeeman transitions vanish. The calculated probabilities of all transitions allowed in the BG regime (1-10) as a function of B-field can be found in \cite{21}. They all smoothly increase with $B$ and reach their maximum values for $B$ of several kG.

\begin{figure}[tb]
	\centerline{\includegraphics[width=.8\columnwidth]{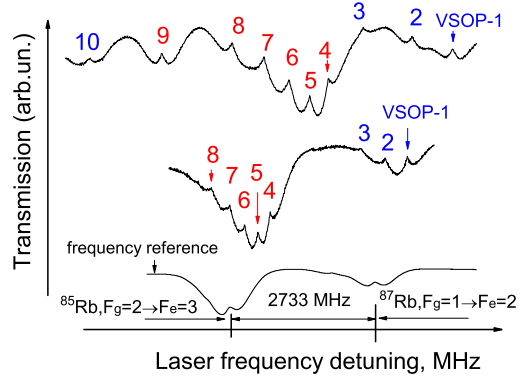}}
	\caption{Rb $D_1$ line SA spectrum in the intermediate fields of 1075~G (upper curve) and 500~G (middle curve). Laser power is 1 mW. The lower curve is the reference.}
	\label{fig:lowfield}
\end{figure}

\begin{figure}[tb]
	\centerline{\includegraphics[width=.8\columnwidth]{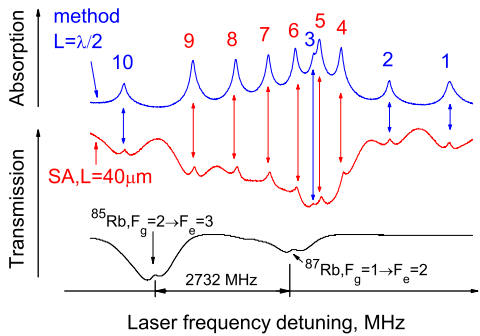}}
	\caption{Comparison of spectra  for $B = 2.2$~kG obtained with the NTC with the thickness $L=\lambda/2$ (upper, blue curve; laser power 50~$\mu$W) and SA spectroscopy (middle, red curve) in the MTC (laser power 1 mW). The lowest curve is the reference.}
	\label{fig:lambda2}
\end{figure}

\begin{figure}[tb]
	\centerline{\includegraphics[width=.8\columnwidth]{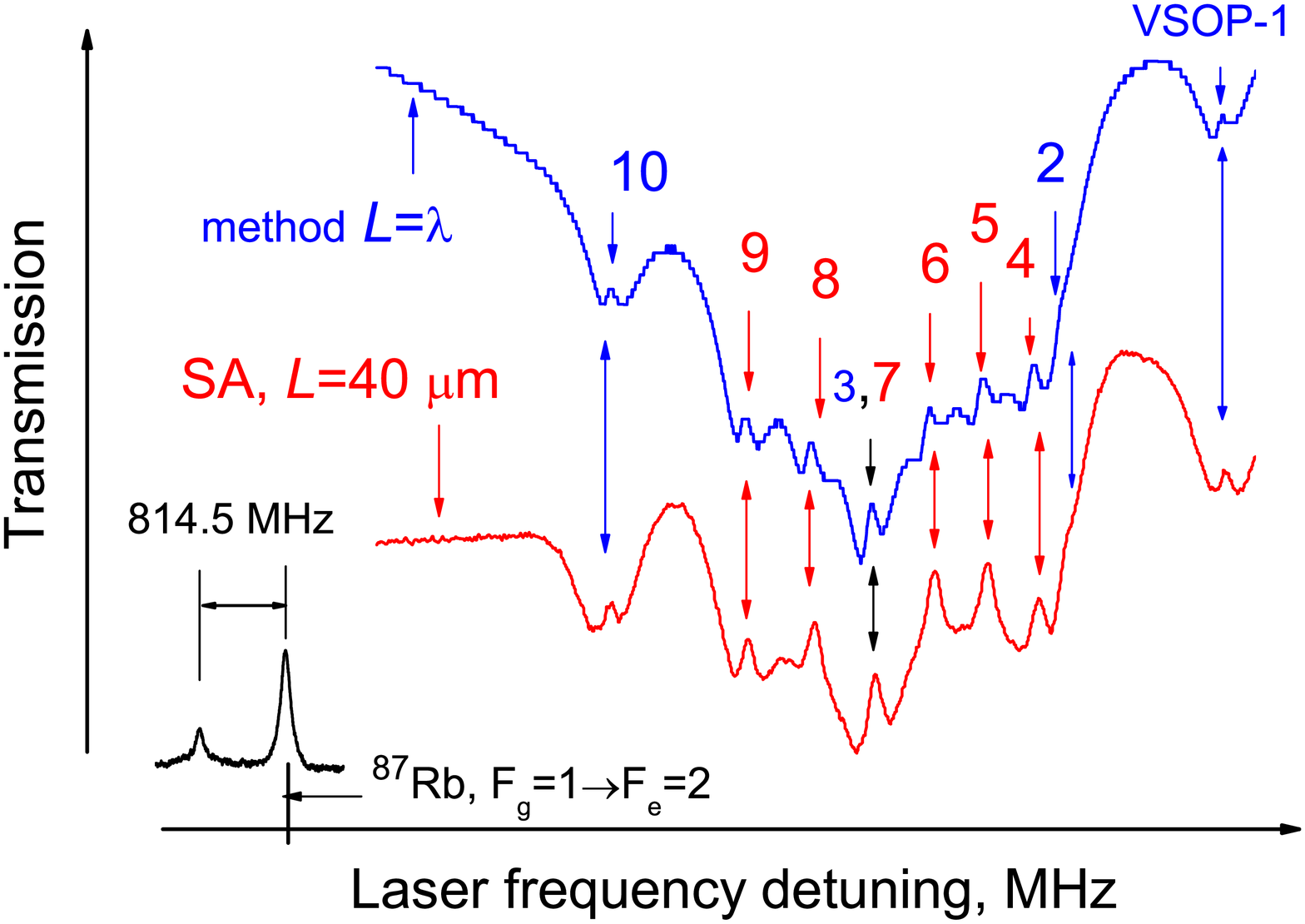}}
	\caption{Comparison of spectra for $B = 5.6$~kG obtained with the NTC with the thickness $L= \lambda$ (upper graph) and SA spectroscopy (middle graph) in the MTC with $L=40\,\mu$m. Laser power is 1 mW. The lowest curve is the reference.
 }
	\label{fig:lambda}
\end{figure}

The SA spectra recorded with MTC in the moderate magnetic fields are shown in Fig. \ref{fig:lowfield}. Atomic transitions labeled 1--3 and 10 for $^{87}$Rb and 4--9 for $^{85}$Rb [see Fig. \ref{fig:setup}(b)] are clearly resolved and no COs are present, thanks to the small thickness of the MTC. Our experiment shows that the presented method is applicable for $B\ge 200$~G, which allows the study of the atomic spectra in the intermediate and high fields, i.e., the nonlinear Zeeman and BG regimes. Examples of high magnetic field SA spectra are shown in Fig. \ref{fig:lambda2} and Fig. \ref{fig:lambda} together with the spectra obtained in the NTC with the thickness $\lambda/2$ and $\lambda$, respectively. Both techniques allow for observation of the high resolution spectra, and we shortly summarize their advantages.

Spectroscopy with the NTC \cite{22, 23} exploits the strong narrowing in absorption spectrum at $L = \lambda$/2 as compared with the case of an ordinary cm-size vapor cell. Particularly, the absorption linewidth for the Rb $D_1$ line reduces to ${\sim} 120$~MHz (FWHM), as opposed to ${\sim} 500$ MHz Doppler width. Moreover, the relative transition probabilities can be directly extracted from the resonance amplitudes, since the spectrum is essentially background-free.
On the other hand, for the NTC with $L=\lambda$, the achieved lines are much narrower (${\sim} 20$~MHz), which is beneficial for investigation of closely-spaced resonances. The peak amplitudes are again proportional to the transition probability as it was demonstrated in [8], however, the presence of the Doppler background makes them harder to extract.

For SA spectroscopy the amplitude ratios of the VSOP do not correspond to the probabilities of the appropriate atomic transitions. Particularly, the ratio of the VSOP peaks 4 and 5 obtained by the $\lambda$-method (Fig.\ref{fig:lambda}) is approximately 1, as calculated in Ref. \cite{21}, whereas the same ratio obtained with the SA method (middle curve) is clearly different. Nevertheless, the observed line widths are narrow, ${\sim} 20$~MHz, and there is a large similarity to the spectra obtained with the $\lambda$-thick NTC.
The great advantage of the SA with the MTC is the simplicity of the setup. Not only it is much easier to manufacture the $40\, \mu$m cell, but also the temperature of its coldest spot (about $70\,^{\circ}$C) is lower than that of the NTC (about $120\,^{\circ}$C). Moreover, the MTC can be fabricated from a standard glass or fused silica \cite{24}, which is resistant to atomic vapor of alkali metals up to $100\,^{\circ}$C for a long time.

\begin{figure}[tb]
	\centerline{\includegraphics[width=.75\columnwidth]{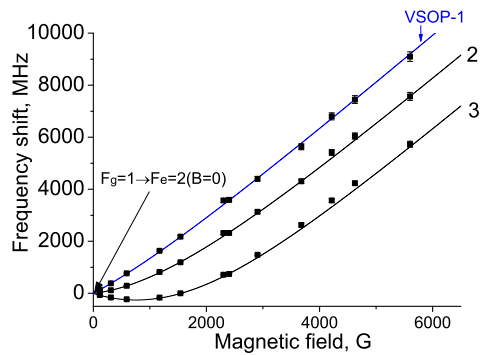}}
	\caption{Frequency shifts of the resonances labeled 1--3 [see Fig. \ref{fig:setup}(b)] in the magnetic field. Solid lines represent the theory and symbols correspond to the experimental results.}
	\label{fig:shift}
\end{figure}

Figure \ref{fig:shift} shows the frequency shift of components 1, 2 and 3 obtained with SA spectroscopy as a function of magnetic field $B$ relative to the initial $F_g$=1 $\rightarrow$ $F_e$=2 position at $B=0$. The magnetic field dependence of the frequency shifts of  transitions 4--9 has been calculated in \cite{21}. For $B \gg B_0$ their frequency slope $s$ asymptotically approaches the same value for transitions 1--10, $s = [g'_J (5P_{1/2}) m'_J  - g_J (5S_{1/2}) m_J] \mu_B/B \approx  1.87$ [MHz/G], where $g_J$ and $g'_J$ are Land\'e factors for the ground and excited levels, respectively. The onset of this value is indicative of the BG regime.
The resonance labeled VSOP-1 is particularly convenient for magnetic field measurements, as it is always located at the high frequency wing of the atomic transition and does not overlap other transitions up to fields of a few T. Its frequency shift can be used to determine the magnetic field strength in the volume defined by the size of the laser beam and the thickness of the MTC. This enables precision field mapping in 1D with a spatial resolution equal to the cell thickness \cite{17}. Moreover, application of the CCD camera instead of a photodiode in the detection system could simultaneously provide the magnetic-field mapping in the other two dimensions.

Long-term stabilization of a laser frequency is routinely achieved by electronic locking of the laser to an appropriate spectroscopic reference signal. For many atomic physics experiments, the light frequency has to be detuned from the exact resonance by some GHz, e.g., to avoid absorption and exploit the dispersive properties of the medium \cite{28} or to precisely control the AC-stark shifts introduced by the detuned beam \cite{29}. In some cases the other isotope lines can be used to stabilize the laser frequency or the beat-note signal of the two lasers can be used for offset lock. The latter, however, requires two lasers, and is practically limited to around 10~GHz detunings \cite{30}. Here we propose a simple atomic frequency reference system which consists of MTC with $L=40\, \mu$m and two permanent ring magnets separated by a plastic spacer. The heating of the MTC is provided by a simple hot air blower. For the 15~mm spacer, the achieved field value is $B=5.6$~kG  and the VSOP-1 resonance for $\sigma^{+}$ excitation is shifted by 9.2~GHz with respect to its $B=0$ position (Fig. \ref{fig:shift}).
For $\sigma^{-}$ excitation a similar resonance can be used in the low frequency wing of the spectrum. Thus, by changing the spacer thickness and light polarization it is possible to form the frequency reference in a wide frequency range, limited by the strength of available magnets. Thanks to the narrowness of the MTC the accuracy of such a reference does not suffer from field inhomogeneities.

\begin{figure}[tb]
	\centerline{\includegraphics[width=.9\columnwidth]{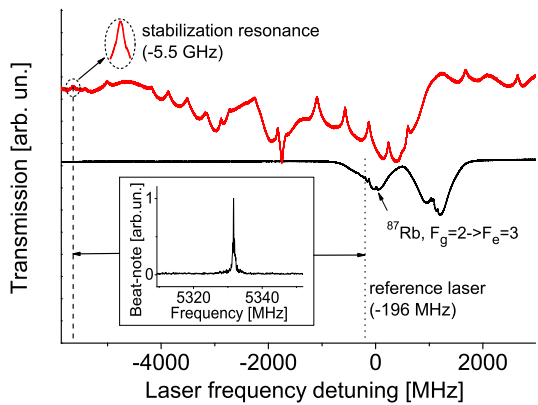}}
	\caption{SA spectrum of the rubidium $D_2$ line in the MTC in $B=2.2$~kG (upper, red curve). Bottom (black) trace is the reference cell spectroscopy in $B=0$. The resonance indicated by the dashed line was used for frequency stabilization of the diode laser. The inset graph shows the beat-note signal against a second, frequency locked (dotted line), laser. }
	\label{fig:beatnote}
\end{figure}

We have verified that although for $B>2$~kG there are twenty transitions in the Rb $D_2$ line (for either $\sigma^{+}$ or $\sigma^{-}$ polarization), the method could be implemented successfully also for that line. Figure \ref{fig:beatnote} illustrates the stabilization of a tunable laser frequency to the magnetically shifted resonance. The beatnote signal against the reference laser, frequency locked $196$~MHz below the $^{87}$Rb $F_g=2\rightarrow F_e=3$ transition, demonstrates the feasibility of over 5 GHz laser detuning to the frequency where no resonance is present at $B=0$.

In conclusion, we have demonstrated successful implementation of SA spectroscopy based on micrometric-thin cell for Doppler-free studies of  atomic transitions in a wide region of magnetic fields, from 0.2 kG up to 6 kG. The is mainly due to elimination of COs in the thin cell and the resulting simplification of the atomic spectra. By using circularly polarized light, the number of atomic transitions which are spectrally resolved is lowered to four (six) $D_1$ transitions of $^{87}$Rb ($^{85}$Rb). The system based on MTC and two permanent magnets is proposed as a tunable atomic frequency reference with a tuning range of the order of $\pm 10$~GHz around the zero-magnetic-field optical transitions frequencies, limited by the currently available magnets. Relatively simple manufacturing of the MTC (as compared with NTC) is advantageous for practical implementation of the proposed method.
The MTCs are universal and can be used for large field spectroscopy or the optical magnetometry with a micrometric spatial resolution, important for mapping strongly inhomogeneous magnetic fields.

This research was supported by a Marie Curie International Research Staff Exchange Scheme Fellowship, FP7  "Coherent optics sensors for medical applications-COSMA" (PIRSES-GA-2012-295264), POIG 02.01.00-12-023/08, Foundation for Polish Science (TEAM Programme), National Science Centre (2012/07/B/ST2/00251), and Jagiellonian University (SET project co-financed by the EU).

\pagebreak

\end{document}